\documentclass[aps,prl,reprint,twocolumn,superscriptaddress]{revtex4-2}

\usepackage{hyperref}
\usepackage{graphicx} 
\usepackage{dcolumn}   
\usepackage{bm}        
\usepackage{amssymb}   
\usepackage{amsfonts}
\usepackage{graphics}
\usepackage{epsfig}
\usepackage{color}

\begin{document}

\title{Heavy-flavor transport and hadronization in pp collisions}

\author{Andrea Beraudo}
\affiliation{INFN - Sezione di Torino, via Pietro Giuria 1, I-10125 Torino}
\email{beraudo@to.infn.it}
\author{Arturo De Pace}
\affiliation{INFN - Sezione di Torino, via Pietro Giuria 1, I-10125 Torino}
\author{Daniel Pablos}
\affiliation{INFN - Sezione di Torino, via Pietro Giuria 1, I-10125 Torino}
\affiliation{Departamento de F\'isica, Universidad de Oviedo, Avda. Federico Garc\'ia Lorca 18, 33007 Oviedo, Spain}
\affiliation{Instituto Universitario de Ciencias y Tecnolog\'ias Espaciales de Asturias (ICTEA), Calle de la Independencia 13, 33004 Oviedo, Spain}
\author{Francesco Prino}
\affiliation{INFN - Sezione di Torino, via Pietro Giuria 1, I-10125 Torino}
\author{Marco Monteno}
\affiliation{INFN - Sezione di Torino, via Pietro Giuria 1, I-10125 Torino}
\author{Marzia Nardi}
\affiliation{INFN - Sezione di Torino, via Pietro Giuria 1, I-10125 Torino}
\date{\today}

\begin{abstract}
{Recent experimental results on the $\Lambda_c^+/D^0$ ratio in proton-proton collisions have revealed a significant enhancement compared to expectations based on universal fragmentation fractions/functions across different colliding systems, from $e^+e^-$ to pp. This unexpected enhancement has sparked speculation about the potential effects of a deconfined medium impacting hadronization, previously considered exclusive to heavy-ion collisions. In this study, we propose a novel approach that assumes the formation of a small, deconfined, and expanding fireball even in pp collisions, where charm quarks can undergo rescattering and hadronization. We make use of the same in-medium hadronization mechanism developed for heavy-ion collisions, which involves local color-neutralization through recombination of charm quarks with nearby opposite color charges from the background fireball. Our model incorporates the presence of diquark excitations in the hot medium, which promotes the formation of charmed baryons. Moreover, the recombination process, involving closely aligned partons from the same fluid cell, effectively transfers the collective flow of the system to the final charmed hadrons. We show that this framework can qualitatively reproduce the observed experimental findings in heavy-flavor particle-yield ratios, $p_T$-spectra and elliptic-flow coefficients. Our results provide new, complementary supporting evidence that the collective phenomena observed in small systems naturally have the same origin as those observed in heavy-ion collisions.}

\end{abstract}

\maketitle

\paragraph*{Introduction.}
Recent heavy-flavor measurements in proton-proton collisions found a surprisingly large value of the $\Lambda_c^+/D^0$ and $\Xi_c^0/D^0$ ratios~\cite{ALICE:2020wfu,ALICE:2021psx}, strongly enhanced with respect to expectations based on fragmentation fractions extracted from $e^+e^-$ data and compatible with the results obtained in heavy-ion collisions. 
{Given that in this last case the baryon enhancement, primarily observed at intermediate transverse-momentum values, is commonly attributed to a recombination process between heavy quarks and thermal partons from the hot, deconfined medium generated after the collision of two nuclei, this raises the question of whether a similar mechanism of hadronization can occur in proton-proton collisions, in which a small droplet of Quark-Gluon Plasma (QGP) might also be produced.}
This was the idea proposed for instance in Refs.~\cite{Minissale:2020bif,Song:2018tpv}, which led the authors to satisfactory describe the $\Lambda_c^+/D^0$ ratio measured in pp collisions at the LHC. Another attempt to interpret the enhanced production of charmed baryons was based on the Statistical Hadronization Model~\cite{He:2019tik}, assuming a thermal population of the different charmed meson and baryon states predicted by the Relativistic Quark Model around a universal hadronization temperature.
Reproducing the above observations is a challenge for QCD event generators, but recent Color-Reconnection (CR) models implemented in PYTHIA 8~\cite{Christiansen:2015yqa} can provide a satisfactory description of the data allowing a rearrangement of the confining potential among the partons before hadronization (with the possible formation of junction topologies), which decreases the potential energy stored in the color field and favors the production of baryons. 

{In this paper we propose that the same mechanism of heavy-flavor hadron production at work in heavy-ion collisions occurs also in the pp case. This entails that also in proton-proton collisions a small deconfined fireball, with a hydrodynamic expansion driven by pressure gradients, is formed. Such a hot medium affects the stochastic propagation, modeled through a relativistic Langevin equation, of the heavy quarks before hadronization and acts as a reservoir of color charges with which they can undergo recombination when reaching a fluid cell around the QCD hadronization temperature $T_H$.}
Both the description of the heavy-quark dynamics through the fireball in terms of a relativistic Langevin equation~\cite{Alberico:2013bza,Beraudo:2017gxw} and the modelling of their local color-neutralization with opposite-charge thermal particles during hadronization~\cite{Beraudo:2022dpz} have been described in detail in previous publications. In the following we will give just a brief summary of the main points, focusing instead on providing a realistic modelling of the bulk background environment produced in proton-proton collisions.

\begin{figure}
\includegraphics[width=0.24\textwidth]{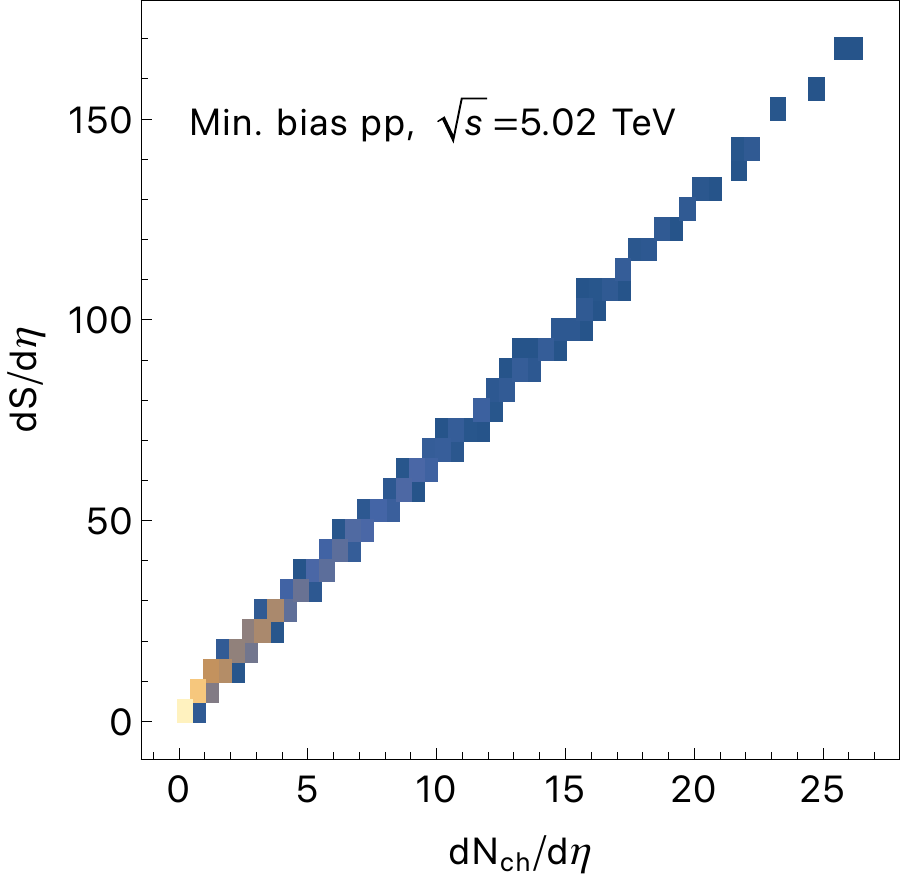}
\includegraphics[width=0.23\textwidth]{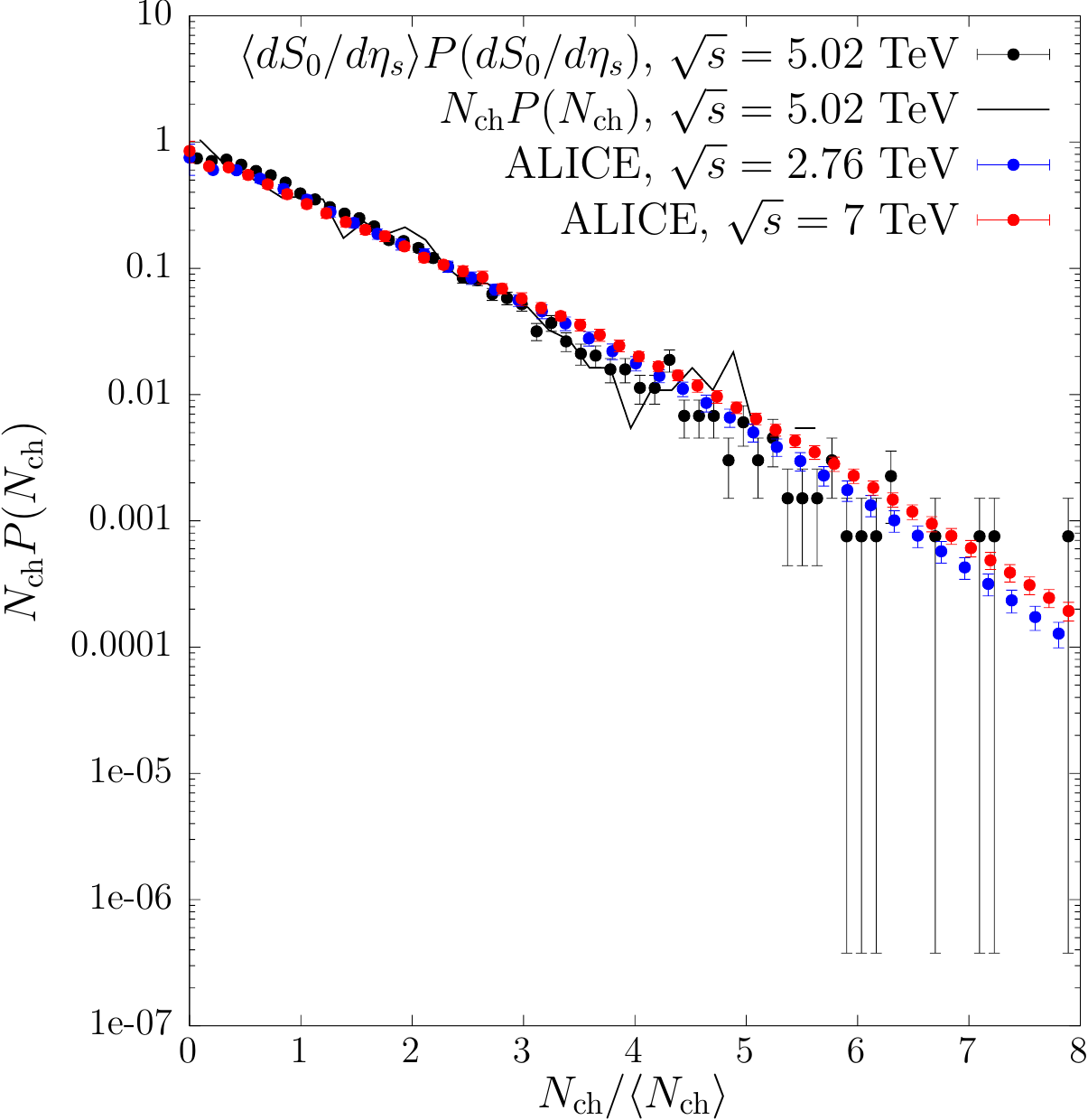}
\caption{Left panel: correlation between the (spacetime/pseudo-)rapidity density of initial deposited entropy and final charged particles in our modelling of pp collisions at $\sqrt{s}\!=\!5.02$ TeV. Right panel: KNO scaling~\cite{Koba:1972ng} of the charged particle distribution in pp collisions. Results referring to the initial conditions and hydrodynamic calculations employed in this work are compared to ALICE data from Ref.~\cite{ALICE:2015olq} for NSD events at $\sqrt{s}=2.76$ TeV (blue dots) and $\sqrt{s}=7$ TeV (red dots).}\label{fig:initial}
\end{figure}

\paragraph*{Theoretical framework.}
{The amount of entropy deposited at midrapidity by the collision of two nucleons can be constrained by analyzing the final particle multiplicity distribution. It is well known that many of the observable features associated to the emergence of collectivity can be understood using the assumption that the deposited entropy behaves hydrodynamically shortly after the collision, translating spatial asymmetries of the initial condition into anisotropies of the momentum distributions of the final particles~\cite{Heinz:2013th,Gale:2013da}. 
In particular, confrontation with experiments involving small colliding systems (i.e. proton-proton and proton-nucleus) has highlighted the role played by sub-nucleonic fluctuations~\cite{Weller:2017tsr}. In order to obtain realistic event-by-event (EBE) initial conditions we use the T$_{\rm R}$ENTo model~\cite{Moreland:2014oya}}, which simulates the initial entropy deposition by the sub-nucleonic constituents of the two protons, here assumed to collide  at $\sqrt{s}\!=\!5.02$ TeV. This requires setting values for a few parameters, related to the dependence of the deposited entropy on the thickness function of the two incoming protons ($p\!=\!0$, corresponding to a geometric mean, {akin to the result obtained from gluon saturation dynamics at small $x$~\cite{Schenke:2012wb}}), to the fluctuating weight of each constituent to the process ($k\!=\!0.3$), to the number of sub-nucleonic constituents ($n_c\!=\!6$) and to the nucleon ($w\!=\!0.92$ fm) and constituent ($v\!=\!0.43$ fm) widths \footnote{All these parameters were calibrated using Bayesian inference~\cite{Moreland:2018gsh}, but we have had to modify $k$ from 0.19 to 0.3 in order to describe the KNO scaling in pp collisions measured in experiments, see right panel of Fig.~\ref{fig:initial}}. An overall normalization is also necessary in order to ensure that at the end of the hydrodynamic evolution one obtains a charged-particle pseudorapidity density $dN_{\rm ch}/d\eta$ (in this description coming entirely from soft processes) in agreement with the measured one. 
At the initial longitudinal proper time $\tau_0$ the entropy density around space-time rapidity $\eta_s\!=\!0$ is given by
\begin{equation}
s_0(\vec x_\perp)|_{\eta_s=0}=\left.\frac{dS_0}{d\vec x_\perp dz}\right|_{z=0}=\frac{1}{\tau_0}\left(\frac{dS_0}{d\vec x_\perp d\eta_s}\right)_{\eta_s=0}\,,
\end{equation}
where the quantity in parentheses is the T$_{\rm R}$ENTo output, while $s_0(\vec x_\perp)$ (in units of fm$^{-3}$) is the quantity used to initialize at $\tau_0\!=\!0.4$ fm/c the subsequent 2+1 relativistic hydrodynamic evolution, evaluated with the  MUSIC   code~\cite{Schenke:2010nt,Schenke:2010rr,Paquet:2015lta}. Hydrodynamic equations are solved using the equation of state computed by the Hot-QCD Collaboration~\cite{HotQCD:2014kol} and setting a constant shear viscosity $\eta/s=0.13$ and non-vanishing bulk viscosity over entropy density parameter $\zeta/s$, whose temperature dependence we parameterize as in~\cite{Denicol:2009am}.  
The deposited entropy around mid-rapidity is then given by:
\begin{equation}
    \frac{dS_0}{d\eta_s}=\tau_0\int d\vec x_\perp s_0(\vec x_\perp)\,.
\end{equation}
The latter is the quantity directly related to final charged-particle multiplicity per unit rapidity. As one can see in the left panel of Fig.~\ref{fig:initial} there is a perfect linear correlation between the two quantities, $dS/d\eta_s\sim dN_{\rm ch}/d\eta$, with proportionality coefficient $K\!\approx\!7.2$. {Integrated particle distributions are obtained via the Cooper-Frye method~\cite{Cooper:1974mv} by particlizing the fluid cells on a isothermal freezeout hypersurface, taking for soft hadrons the decoupling temperature $T_{\rm FO}\!=\!145$ MeV.} Our initial conditions, after EBE hydrodynamic evolution, provide an average $\langle dS/d\eta_s\rangle\!=\!37.59$ for minimum-bias pp collisions, in good agreement with the estimate found in Ref.~\cite{Hanus:2019fnc}. This translates into a final charged-particle multiplicity $dN_{\rm ch}/d\!\eta\!\approx 5.22$, to be compared with the experimental values $4.63^{+0.30}_{-0.19}$ and $5.74^{+0.15}_{-0.15}$ measured  by ALICE~\cite{ALICE:2015olq} in non single-diffractive proton-proton collisions at $\sqrt{s}\!=\!2.76$ and 7 TeV, respectively. Besides verifying that the present initialization and hydrodynamic evolution lead to the correct average hadron multiplicity, in the right panel of Fig.~\ref{fig:initial} one can see that the charged particle multiplicity ($N_{\rm ch}$) distribution itself, with its KNO scaling~\cite{Koba:1972ng}, is reasonably well described. Thus, by selecting the 0-1\% percentile of the initial $dN_{\rm ev}/dS_0$ distribution we also construct a sample of about $10^3$ high-multiplicity events, with $\langle dS/d\eta_s\rangle\!=\!187.53$, storing the information on their hydrodynamic evolution.
Although our purpose is not to perform a precision study of soft observables,
these checks allowed us to validate the model of the fireball assumed to be produced in proton-proton collisions against experimental data. This is crucial to ensure that we get a realistic description of the background medium in which the heavy quarks propagate and undergo hadronization.

\begin{figure}
\includegraphics[width=0.25\textwidth]{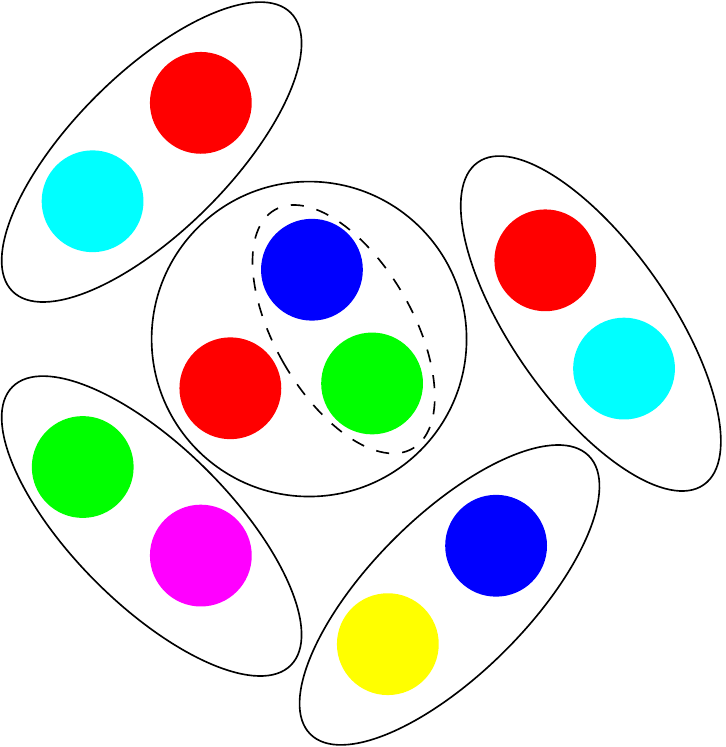}
\caption{A cartoon of our \emph{local} color-neutralization mechanism via Quark-antiquark or Quark-diquark recombination of opposite charges within the same fluid cell.}\label{fig:cartoon}
\end{figure}
The initial quark-antiquark pairs are generated with the POWHEG-BOX~\cite{Alioli:2010xd} tool, which simulates the hard event and -- once interfaced with PYTHIA -- the associated initial and final-state parton shower and other non-perturbative effect such as the intrinsic $k_T$ of the incoming partons. In this paper we focus on charm, for which more experimental data are available. In the generation of the hard events we set the charm quark mass to $m_c\!=\!1.3$ GeV and we employ the default factorization and renormalization scales. Both in the minimum-bias and high-multiplicity cases -- each sample of initial conditions for the fireball evolution containing about $10^3$ independent proton-proton collisions -- we generate $10^7$ $c\overline c$ pairs, which are distributed among the different pp events according to their initial $dS/d\eta_s$. The initial position of the pairs in the transverse plane is then sampled according to the local entropy density $s_0(\vec x_\perp)$. Hence, even in the minimum-bias sample, the $c\overline c$ pairs tend to be concentrated in the hot spots of the events with the largest $dS/d\eta_s$. This is the analogous of the so-called ``pedestal effect''~\cite{UA1:1983hhd}, i.e. the fact that hadronic collisions containing pairs of jets are characterized by a higher activity also outside the jet cone. As a result, when distributing the $c\overline c$ pairs among the different pp events of the minimum-bias sample, only about $5\%$ of them are found at $\tau_0$ in a fluid cell below the hadronization temperature $T_H\!=\!155$ MeV; this fraction drops to $1\%$ when considering the high-multiplicity sample. This is crucial to understand why the same modifications of the heavy-flavor hadrochemistry supposed to be a distinctive feature of nuclear collisions are also observed in the pp case: the shorter lifetime of the fireball going from AA to pp collisions affects only the kinematic distributions of the final hadrons arising from the recombination process, but not their integrated yields, which simply depend on the existence of a color-reservoir and not on its collective flow.

Starting from the longitudinal proper time $\tau_0$ we simulate the stochastic dynamics of charm quarks through the fireball. No pre-equilibrium evolution neither of the heavy quarks nor of the medium is considered. The heavy-quark propagation in the expanding QGP is described through the relativistic Langevin equation. The latter provides a recipe to update the heavy-quark momentum during the time-step $\Delta t$ in the local rest-frame (LRF) of the fluid:
\begin{equation}
   {\Delta \vec{p}}/{\Delta t}=-{\eta_D(p)\vec{p}}+{\vec\xi(t)},\label{eq:Langevin} 
\end{equation}
where $\vec \xi$ is a noise term responsible for the in-medium momentum broadening of the particle, specified by its temporal correlator
$\langle\xi^i(\vec p_t)\xi^j(\vec p_{t'})\rangle\!=\!{b^{ij}(\vec p_t)}{\delta_{tt'}}/{\Delta t}$, with
\begin{equation}
 {b^{ij}(\vec p)}\!\equiv\!{\kappa_\|(p)} \hat{p}^i\hat{p}^j+{\kappa_\perp(p)}(\delta^{ij}\!-\!\hat p^i\hat p^j).   
\end{equation}
In the above the transport coefficients $\kappa_{\|/\perp}$ quantify the average longitudinal/transverse squared momentum exchange per unit time with the medium. In the following for them we use results provided by weak-coupling (Hard-Thermal-Loop, HTL) and the most recent lattice-QCD (lQCD) calculations~\cite{Altenkort:2023oms}. Once the $\kappa_{\|/\perp}$ are known, the friction coefficient $\eta_D$ is fixed by a generalized Einstein relation in order to ensure the approach to kinetic equilibrium. More details can be found in~\cite{Alberico:2013bza,Beraudo:2017gxw}. It is interesting to quantify the time spent by the heavy quarks in the deconfined fireball before hadronization. In our sample of minimum-bias pp collisions for the average longitudinal proper time at which charm quarks undergo hadronization we found $\langle\tau_H\rangle\!\approx\! 1.95$ fm/c, with a $\tau_H^{\rm max}\approx 4.25$ fm/c. In the $0\!-\!1\%$ high-multiplicity sample we found $\langle\tau_H\rangle\!\approx\! 2.92$ fm/c, with a $\tau_H^{\rm max}\approx 4.79$ fm/c. As we will see, the longer time spent in a fireball with larger pressure gradients will lead to a larger radial flow of charmed hadrons in high-multiplicity pp collisions.

\begin{figure*}
\includegraphics[width=0.9\textwidth]{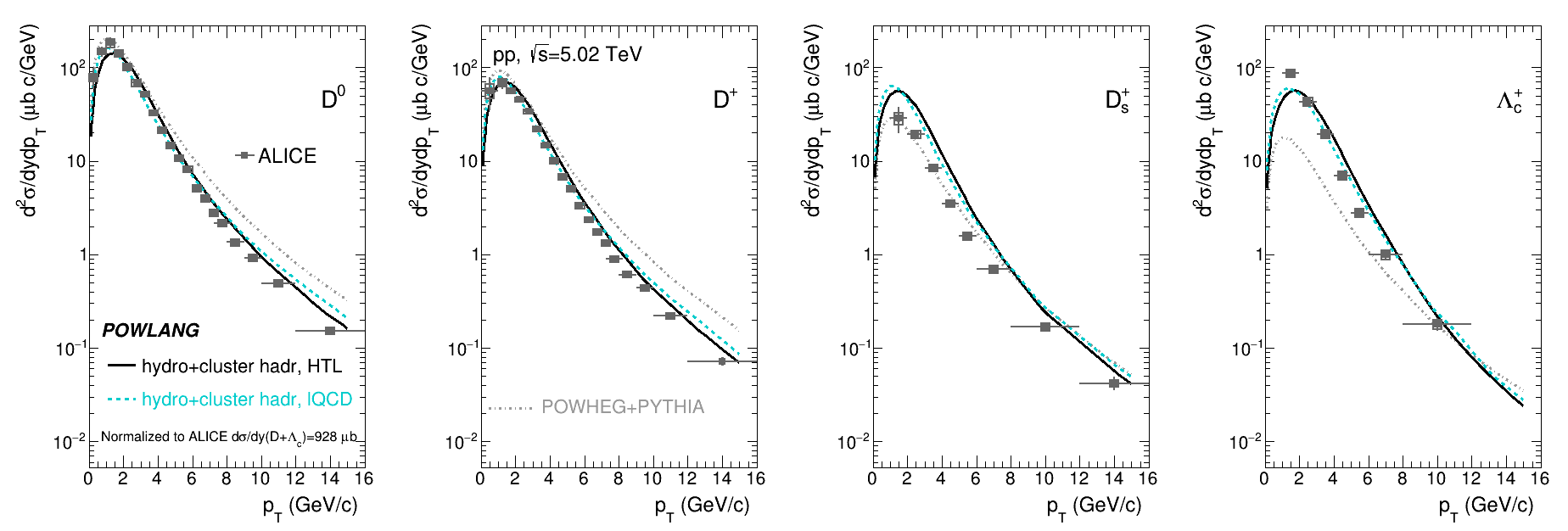}
\caption{Charmed hadron $p_T$-distributions in pp collisions at $\sqrt{s}\!=\!5.02$ TeV normalized to the ALICE estimate for the $D+\Lambda_c^+$ cross-section. Results obtained with POWHEG-BOX standalone (dotted grey curves) and supplemented with an in-medium transport+hadronization stage with HTL (continuous black curves) and lattice-QCD (dashed cyan curves) are compared to ALICE data~\cite{ALICE:2020wfu,ALICE:2021mgk}.}\label{fig:ptspectrum}
\end{figure*}
We now briefly summarize the hadronization model employed in this paper, based on a \emph{local} color neutralization mechanism discussed in detail in~\cite{Beraudo:2022dpz} and illustrated in Fig.~\ref{fig:cartoon}. We assume that once a $c$ quark reaches the hadronization hypersurface at $T_H\!=\!155$ MeV it undergoes recombination with an opposite color charge -- either a light antiquark or a diquark, both assumed to populate the fireball around $T_H$ with their respective thermal abundance -- \emph{from the same fluid cell}: long-range interactions are in fact screened by the medium and, furthermore, this choice leads to a minimization of the confining potential. Diquark masses are the ones employed in PYTHIA 6.4~\cite{Sjostrand:2006za}. For simplicity, no $\gamma_s$ fugacity factor is introduced to suppress strange quarks in low-multiplicity pp events, which in any case provide a minor contribution to charm production. Even in the minimum-bias sample, if one weights each proton-proton collision by the average number of $c\overline c$ pairs produced in that event one would get $\langle dS/d\eta_s\rangle_{c\overline c}\!\approx\!68.87$, corresponding to $dN_{\rm ch}/d\eta\!\approx\!9.56$, a multiplicity at which the observed enhancement of strange particle production is already substantial~\cite{ALICE:2016fzo}.  After selecting from a thermal distribution both the species and the momentum of the heavy-quark companion in the LRF of the fluid, a color-singlet cluster is constructed. Since the recombinig partons belong to the same fluid cell, in the laboratory frame there is a strong correlation -- referred to as Space-Momentum Correlation (SMC) -- between the heavy-quark position, its momentum and the one of its light companion. Hence recombination usually occurs between quite collinear particles in the laboratory frame: this favors the formation of low invariant-mass clusters. As in the HERWIG event generator~\cite{Webber:1983if}, light clusters (below an invariant mass around 4 GeV) undergo a two-body decay, producing a charmed hadron accompanied by a pion ($65\%$ of cases) or a photon, if only this last channel is kinematically open or if it is predicted by the PDG for resonances around that invariant mass. The decay is isotropic in the cluster rest-frame, but due to SMC the cluster is a boosted object along the direction of expansion of the fireball. Hence this local recombination scheme is an efficient mechanism to transfer the collective flow of the fireball to the final charmed hadrons. Concerning higher invariant-mass clusters, these are treated as strings and hadronized with PYTHIA 6.4~\cite{Sjostrand:2006za}, which simulates their fragmentation into the final hadrons. The present local hadronization mechanism is quite schematic, but it has three main virtues: at variance with a pure $2/3\to 1$ coalescence process, thanks to its $2\to 1^*\to N$ dynamics, it conserves four-momentum exactly; it accounts for the enhanced baryon production via recombination with diquarks; it includes by construction SMC, which has a deep impact on the momentum and angular distributions of the final particles~\cite{Beraudo:2022dpz}.
Notice, finally, that the request of recombining with unit probability the heavy quark with the nearest opposite color-charge is not so strong as it may appear, since the process does not entail the direct formation of a bound state, with the necessity of convoluting the parent quark distributions with the Wigner function of the final hadron, as in coalescence approaches.

\paragraph*{Results.}
The predictions of our model for the charmed-hadron $p_T$-differential production cross-sections in proton-proton collisions are shown in Fig.~\ref{fig:ptspectrum} for $D^0$, $D^+$ and $D_s^+$ mesons, and for $\Lambda_c^+$ baryons.
Indeed, current pQCD predictions tend to underestimate the $c\overline c$ production cross-section in hadronic collisions, which lies on the upper edge of the theoretical uncertainty band~\cite{ALICE:2021dhb}. Since we are mainly interested into the relative yields of the various particles and on the shape of their momentum distributions, in Fig.~\ref{fig:ptspectrum} we display the $p_T$-spectra of the different charmed hadrons rescaled to a common normalization given by the experimental $D+\Lambda_c$ production cross-section measured by ALICE~\cite{ALICE:2021dhb}. In the different panels we display the results obtained with POWHEG-BOX standalone -- in which, beside the parton-shower, also the hadronization stage is simulated with PYTHIA 6.4 -- and the ones in which the formation of a small fireball is assumed, where the heavy quarks undergo rescattering and hadronization according to the previously described model. One can see that POWHEG-BOX standalone underpredicts the $\Lambda_c^+$ production and misses the slope of the spectra. On the other hand, including heavy quark rescattering and hadronization in the fireball leads to a better description of the experimental data, with an enhanced $\Lambda_c^+$ production and steeper $p_T$-spectra.

\begin{figure*}
\includegraphics[width=0.9\textwidth]{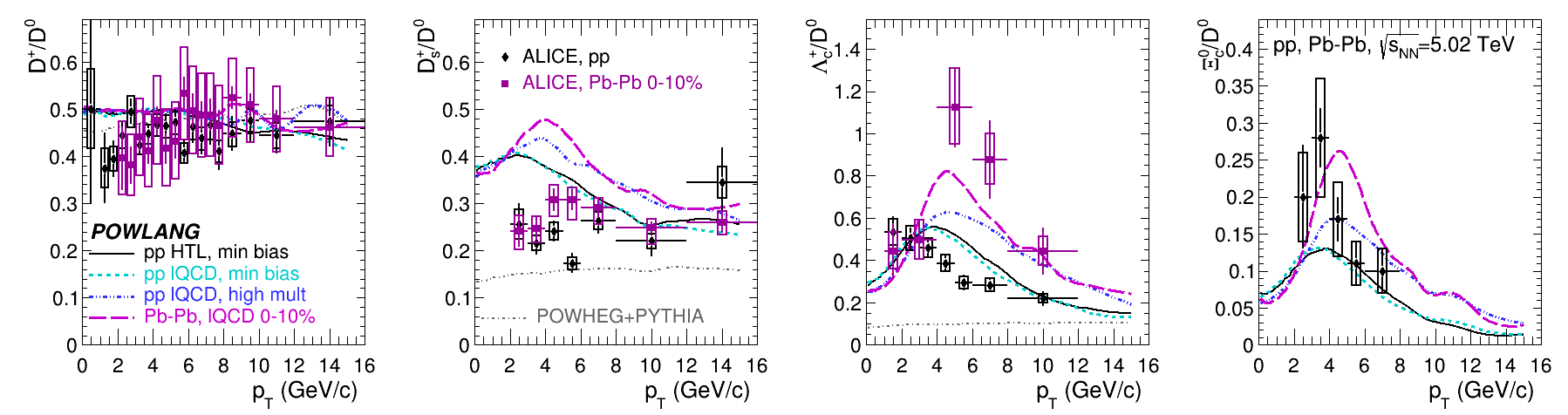}
\caption{Charmed-hadron yield ratios as a function of $p_T$ for different colliding systems at $\sqrt{s_{\rm NN}}\!=\!5.02$ TeV. Predictions including in-medium transport+hadronization in minimum-bias and high-multiplicity pp collisions and in central PbPb collisions (see legend) are compared to ALICE data~\cite{ALICE:2020wfu,ALICE:2021mgk,ALICE:2021rxa,ALICE:2021bib,ALICE:2021psx}. The enhanced baryon-to-meson ratio and the shift of its peak in denser systems is qualitatively well reproduced. Also shown are the pp predictions of POWHEG+PYTHIA standalone, with no medium effects, which undershoots charmed baryon production.}\label{fig:ratios}
\end{figure*}

In Fig.~\ref{fig:ratios} we plot, as a function of $p_T$, various charmed-hadron yield ratios relative to the one of $D^0$ mesons. One can appreciate the enhanced $D_s^+$ and charmed-baryon production with respect to expectations based on vacuum fragmentation: this experimental observation is reproduced by our model (which tends to slightly overestimate $D_s^+$ production).
The most striking feature of the data, well described by our model, is the peak in the baryon/meson ratio for $p_T\!\approx\! 3\!-\!5$ GeV/c arising from the radial flow of light diquarks. The peak moves to higher values of $p_T$ going from minimum-bias to high-multiplicity pp and, eventually, to central Pb-Pb collisions, consistently with the higher average radial velocity of the fluid cells where heavy-quark hadronization occurs ($\langle u_\perp\rangle_{\rm pp}^{\rm mb}\!\approx\! 0.33\!-\!0.34$, $\langle u_\perp\rangle_{\rm pp}^{\rm hm}\!\approx\! 0.53\!-\!0.54$ and $\langle u_\perp\rangle_{\rm PbPb}^{0-10\%}\!\approx\!0.65\!-\!0.67$, respectively, depending on the choice of the transport coefficients). Notice that, at very high $p_T$, one naturally recovers the $\Lambda_c^+/D^0\!\approx\!0.1$ result typical of vacuum fragmentation. In fact, in this kinematic domain, charmed hadrons mainly come from the decay of high invariant-mass strings, which fragment as in vacuum.

\begin{figure}
\includegraphics[width=0.35\textwidth]{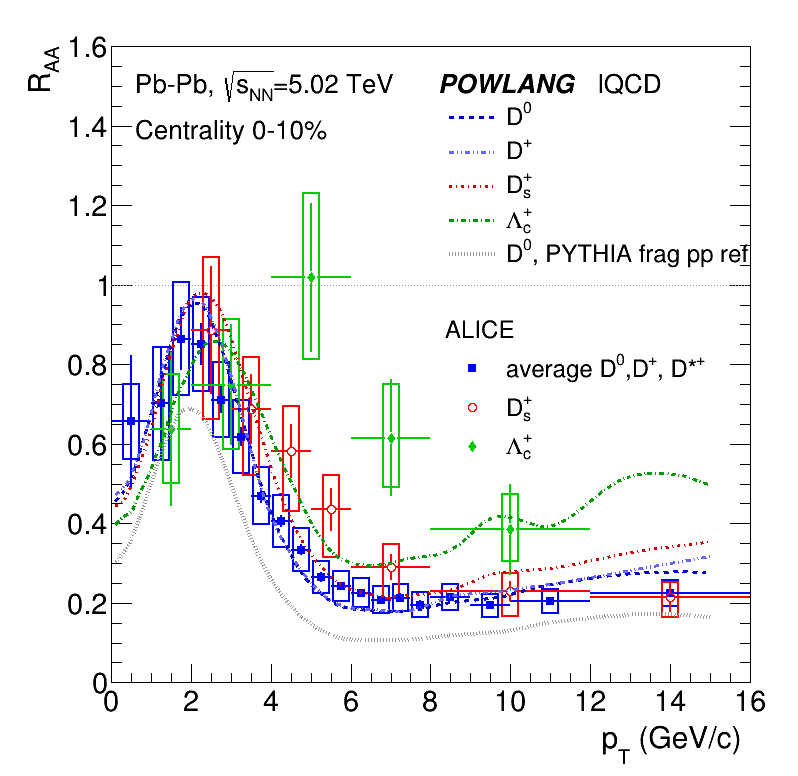}
\caption{Nuclear modification factor for the different charmed hadrons. Theory curves, obtained with lQCD transport coefficients, includes in-medium transport+hadronization in the minimum-bias pp benchmark and are compared to recent ALICE data~\cite{ALICE:2021rxa,ALICE:2021bib}. Also shown is the $D^0$ result with no medium effect in the pp benchmark}\label{fig:RAA}
\end{figure}
Our study is also relevant to correctly quantify medium effects in heavy-ion collisions, where the pp benchmark enters for instance in defining the nuclear modification factor $R_{\rm AA}(p_T)\!\propto\!(dN/dp_T)_{\rm AA}/(dN/dp_T)_{\rm pp}$. 
As one can see in Fig.~\ref{fig:RAA}, the inclusion of medium effects in pp collisions allows one to correctly reproduce the location and magnitude of the radial-flow peak (i.e. the reshuffling of the particle momenta, moving from low to moderate $p_T$) and to obtain a species dependence of the results with the same qualitative trend of the experimental data.

\begin{figure}
\includegraphics[width=0.35\textwidth]{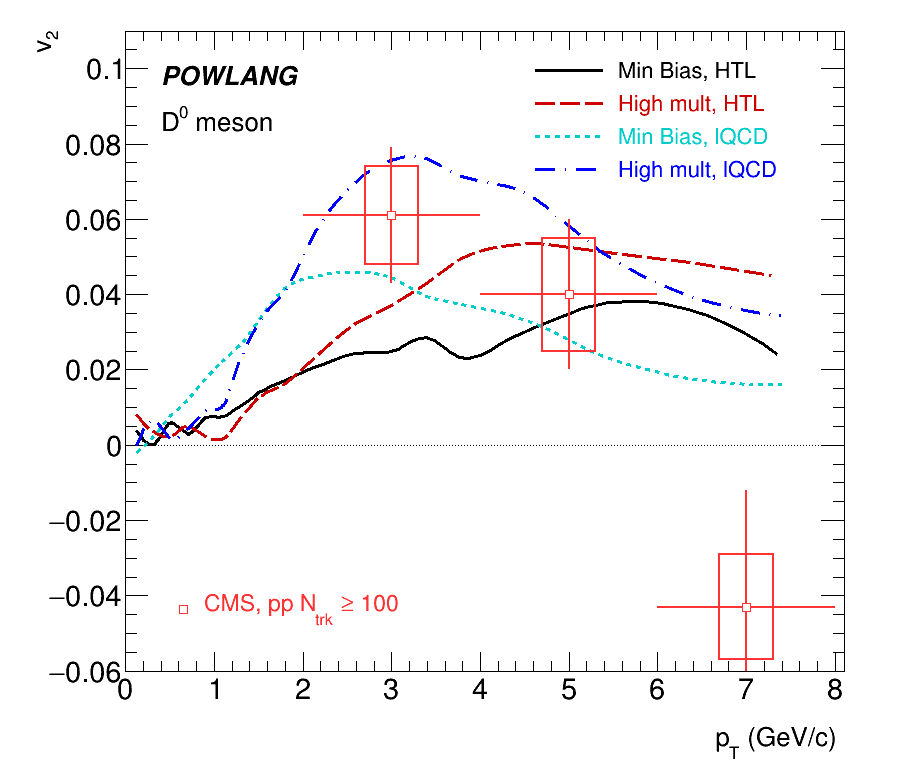}
\caption{Charmed-hadron elliptic-flow coefficient in minimum-bias and high-multiplicity pp collisions at $\sqrt{s}\!=\!5.02$ TeV. Our predictions are compared to CMS results for high-multiplicity pp collisions at $\sqrt{s}\!=\!13$ TeV~\cite{CMS:2020qul}}.\label{fig:v2}
\end{figure}
Finally, it is interesting to study the response of the charmed-hadron azimuthal distributions to the initial deformation of the fireball created in pp collisions, quantified on an EBE basis, according to Ref.~\cite{Qiu:2011iv}, via the eccentricity $\epsilon_2$ and the orientation $\psi_2$ of the minor axis of the approximate elliptic distribution of deposited entropy. For the former one ges $\langle\epsilon_2\rangle\!\approx\!0.31$, quite independent from the event activity. One then evaluates the charmed-hadron elliptic-flow coefficient $v_2\!\equiv\!\langle\cos[2(\phi\!-\!\psi_2)]\rangle$, which is plotted in Fig.~\ref{fig:v2} for minimum-bias and high-multiplicity events and compared to CMS data for high-multiplicity pp collisions~\cite{CMS:2020qul}. An important fraction of the $v_2$ is acquired at hadronization and the larger value obtained in high-multiplicity events has to be attributed not to a different initial deformation of the fireball, but to its longer lifetime.

\paragraph*{Conclusions and perspectives.} In summary, the assumption of the formation of a small deconfined fireball also in pp collisions, affecting the propagation and the hadronization of charm quarks, allows one to provide a consistent picture of several experimental observations involving heavy-flavor hadrons: the slope of their momentum distributions, the $p_T$-dependent enhancement of the baryon-to-meson ratios, the nuclear-modification-factor of their $p_T$-distributions in Pb-Pb collisions and the non-vanishing elliptic-flow coefficient observed also in pp collisions.
A more systematic study -- exploring for instance the sensitivity of our results to the effective light quark and diquark masses, different implementations of sub-nucleonic fluctuations in the initial conditions, pre-equilibrium dynamics or changes in the hydrodynamization time $\tau_0$ -- is surely welcome in the near future, but we believe that our major findings will not change, since they are simply based on a local parton-recombination process occurring at a temperature around the confinement crossover within a fluid-cell undergoing a collective flow.
Our results provide independent, strong indications that the collective phenomena observed in small systems naturally have the same origin as those measured in heavy-ion collisions. So far this was only inferred from the study of soft observables, i.e. from light hadrons emitted in the late stage of the fireball evolution. Hence, having shown that the formation of a small QGP droplet in proton-proton collisions can also affect the production of hard particles like heavy-flavor hadrons represents an important phenomenological and conceptual achievement, which, furthermore, entails reconsidering the universality of hadronization. 

As a next step, we plan to extend our study to proton-nucleus collisions and to the hadronization of bottom quarks, so to provide a unified picture of heavy-flavor production across all colliding systems.

\begin{acknowledgments} 
\paragraph*{Acknowledgements.}
D.P. has received funding from the European Union’s Horizon 2020 research and innovation program under the Marie Sklodowska-Curie grant agreement No. 754496. 
\end{acknowledgments}

\bibliographystyle{apsrev4-2}
\bibliography{main}

\begin{thebibliography}{37}%
\makeatletter
\providecommand \@ifxundefined [1]{%
 \@ifx{#1\undefined}
}%
\providecommand \@ifnum [1]{%
 \ifnum #1\expandafter \@firstoftwo
 \else \expandafter \@secondoftwo
 \fi
}%
\providecommand \@ifx [1]{%
 \ifx #1\expandafter \@firstoftwo
 \else \expandafter \@secondoftwo
 \fi
}%
\providecommand \natexlab [1]{#1}%
\providecommand \enquote  [1]{``#1''}%
\providecommand \bibnamefont  [1]{#1}%
\providecommand \bibfnamefont [1]{#1}%
\providecommand \citenamefont [1]{#1}%
\providecommand \href@noop [0]{\@secondoftwo}%
\providecommand \href [0]{\begingroup \@sanitize@url \@href}%
\providecommand \@href[1]{\@@startlink{#1}\@@href}%
\providecommand \@@href[1]{\endgroup#1\@@endlink}%
\providecommand \@sanitize@url [0]{\catcode `\\12\catcode `\$12\catcode
  `\&12\catcode `\#12\catcode `\^12\catcode `\_12\catcode `\%12\relax}%
\providecommand \@@startlink[1]{}%
\providecommand \@@endlink[0]{}%
\providecommand \url  [0]{\begingroup\@sanitize@url \@url }%
\providecommand \@url [1]{\endgroup\@href {#1}{\urlprefix }}%
\providecommand \urlprefix  [0]{URL }%
\providecommand \Eprint [0]{\href }%
\providecommand \doibase [0]{https://doi.org/}%
\providecommand \selectlanguage [0]{\@gobble}%
\providecommand \bibinfo  [0]{\@secondoftwo}%
\providecommand \bibfield  [0]{\@secondoftwo}%
\providecommand \translation [1]{[#1]}%
\providecommand \BibitemOpen [0]{}%
\providecommand \bibitemStop [0]{}%
\providecommand \bibitemNoStop [0]{.\EOS\space}%
\providecommand \EOS [0]{\spacefactor3000\relax}%
\providecommand \BibitemShut  [1]{\csname bibitem#1\endcsname}%
\let\auto@bib@innerbib\@empty
\bibitem [{\citenamefont {Acharya}\ \emph
  {et~al.}(2021{\natexlab{a}})\citenamefont {Acharya} \emph
  {et~al.}}]{ALICE:2020wfu}%
  \BibitemOpen
  \bibfield  {author} {\bibinfo {author} {\bibfnamefont {S.}~\bibnamefont
  {Acharya}} \emph {et~al.} (\bibinfo {collaboration} {ALICE}),\ }\href
  {https://doi.org/10.1103/PhysRevLett.127.202301} {\bibfield  {journal}
  {\bibinfo  {journal} {Phys. Rev. Lett.}\ }\textbf {\bibinfo {volume} {127}},\
  \bibinfo {pages} {202301} (\bibinfo {year} {2021}{\natexlab{a}})},\ \Eprint
  {https://arxiv.org/abs/2011.06078} {arXiv:2011.06078 [nucl-ex]} \BibitemShut
  {NoStop}%
\bibitem [{\citenamefont {Acharya}\ \emph
  {et~al.}(2021{\natexlab{b}})\citenamefont {Acharya} \emph
  {et~al.}}]{ALICE:2021psx}%
  \BibitemOpen
  \bibfield  {author} {\bibinfo {author} {\bibfnamefont {S.}~\bibnamefont
  {Acharya}} \emph {et~al.} (\bibinfo {collaboration} {ALICE}),\ }\href
  {https://doi.org/10.1007/JHEP10(2021)159} {\bibfield  {journal} {\bibinfo
  {journal} {JHEP}\ }\textbf {\bibinfo {volume} {10}},\ \bibinfo {pages}
  {159}},\ \Eprint {https://arxiv.org/abs/2105.05616} {arXiv:2105.05616
  [nucl-ex]} \BibitemShut {NoStop}%
\bibitem [{\citenamefont {Minissale}\ \emph {et~al.}(2021)\citenamefont
  {Minissale}, \citenamefont {Plumari},\ and\ \citenamefont
  {Greco}}]{Minissale:2020bif}%
  \BibitemOpen
  \bibfield  {author} {\bibinfo {author} {\bibfnamefont {V.}~\bibnamefont
  {Minissale}}, \bibinfo {author} {\bibfnamefont {S.}~\bibnamefont {Plumari}},\
  and\ \bibinfo {author} {\bibfnamefont {V.}~\bibnamefont {Greco}},\ }\href
  {https://doi.org/10.1016/j.physletb.2021.136622} {\bibfield  {journal}
  {\bibinfo  {journal} {Phys. Lett. B}\ }\textbf {\bibinfo {volume} {821}},\
  \bibinfo {pages} {136622} (\bibinfo {year} {2021})},\ \Eprint
  {https://arxiv.org/abs/2012.12001} {arXiv:2012.12001 [hep-ph]} \BibitemShut
  {NoStop}%
\bibitem [{\citenamefont {Song}\ \emph {et~al.}(2018)\citenamefont {Song},
  \citenamefont {Li},\ and\ \citenamefont {Shao}}]{Song:2018tpv}%
  \BibitemOpen
  \bibfield  {author} {\bibinfo {author} {\bibfnamefont {J.}~\bibnamefont
  {Song}}, \bibinfo {author} {\bibfnamefont {H.-h.}\ \bibnamefont {Li}},\ and\
  \bibinfo {author} {\bibfnamefont {F.-l.}\ \bibnamefont {Shao}},\ }\href
  {https://doi.org/10.1140/epjc/s10052-018-5817-x} {\bibfield  {journal}
  {\bibinfo  {journal} {Eur. Phys. J. C}\ }\textbf {\bibinfo {volume} {78}},\
  \bibinfo {pages} {344} (\bibinfo {year} {2018})},\ \Eprint
  {https://arxiv.org/abs/1801.09402} {arXiv:1801.09402 [hep-ph]} \BibitemShut
  {NoStop}%
\bibitem [{\citenamefont {He}\ and\ \citenamefont {Rapp}(2019)}]{He:2019tik}%
  \BibitemOpen
  \bibfield  {author} {\bibinfo {author} {\bibfnamefont {M.}~\bibnamefont
  {He}}\ and\ \bibinfo {author} {\bibfnamefont {R.}~\bibnamefont {Rapp}},\
  }\href {https://doi.org/10.1016/j.physletb.2019.06.004} {\bibfield  {journal}
  {\bibinfo  {journal} {Phys. Lett. B}\ }\textbf {\bibinfo {volume} {795}},\
  \bibinfo {pages} {117} (\bibinfo {year} {2019})},\ \Eprint
  {https://arxiv.org/abs/1902.08889} {arXiv:1902.08889 [nucl-th]} \BibitemShut
  {NoStop}%
\bibitem [{\citenamefont {Christiansen}\ and\ \citenamefont
  {Skands}(2015)}]{Christiansen:2015yqa}%
  \BibitemOpen
  \bibfield  {author} {\bibinfo {author} {\bibfnamefont {J.~R.}\ \bibnamefont
  {Christiansen}}\ and\ \bibinfo {author} {\bibfnamefont {P.~Z.}\ \bibnamefont
  {Skands}},\ }\href {https://doi.org/10.1007/JHEP08(2015)003} {\bibfield
  {journal} {\bibinfo  {journal} {JHEP}\ }\textbf {\bibinfo {volume} {08}},\
  \bibinfo {pages} {003}},\ \Eprint {https://arxiv.org/abs/1505.01681}
  {arXiv:1505.01681 [hep-ph]} \BibitemShut {NoStop}%
\bibitem [{\citenamefont {Alberico}\ \emph {et~al.}(2013)\citenamefont
  {Alberico}, \citenamefont {Beraudo}, \citenamefont {De~Pace}, \citenamefont
  {Molinari}, \citenamefont {Monteno}, \citenamefont {Nardi}, \citenamefont
  {Prino},\ and\ \citenamefont {Sitta}}]{Alberico:2013bza}%
  \BibitemOpen
  \bibfield  {author} {\bibinfo {author} {\bibfnamefont {W.~M.}\ \bibnamefont
  {Alberico}}, \bibinfo {author} {\bibfnamefont {A.}~\bibnamefont {Beraudo}},
  \bibinfo {author} {\bibfnamefont {A.}~\bibnamefont {De~Pace}}, \bibinfo
  {author} {\bibfnamefont {A.}~\bibnamefont {Molinari}}, \bibinfo {author}
  {\bibfnamefont {M.}~\bibnamefont {Monteno}}, \bibinfo {author} {\bibfnamefont
  {M.}~\bibnamefont {Nardi}}, \bibinfo {author} {\bibfnamefont
  {F.}~\bibnamefont {Prino}},\ and\ \bibinfo {author} {\bibfnamefont
  {M.}~\bibnamefont {Sitta}},\ }\href
  {https://doi.org/10.1140/epjc/s10052-013-2481-z} {\bibfield  {journal}
  {\bibinfo  {journal} {Eur. Phys. J. C}\ }\textbf {\bibinfo {volume} {73}},\
  \bibinfo {pages} {2481} (\bibinfo {year} {2013})},\ \Eprint
  {https://arxiv.org/abs/1305.7421} {arXiv:1305.7421 [hep-ph]} \BibitemShut
  {NoStop}%
\bibitem [{\citenamefont {Beraudo}\ \emph {et~al.}(2018)\citenamefont
  {Beraudo}, \citenamefont {De~Pace}, \citenamefont {Monteno}, \citenamefont
  {Nardi},\ and\ \citenamefont {Prino}}]{Beraudo:2017gxw}%
  \BibitemOpen
  \bibfield  {author} {\bibinfo {author} {\bibfnamefont {A.}~\bibnamefont
  {Beraudo}}, \bibinfo {author} {\bibfnamefont {A.}~\bibnamefont {De~Pace}},
  \bibinfo {author} {\bibfnamefont {M.}~\bibnamefont {Monteno}}, \bibinfo
  {author} {\bibfnamefont {M.}~\bibnamefont {Nardi}},\ and\ \bibinfo {author}
  {\bibfnamefont {F.}~\bibnamefont {Prino}},\ }\href
  {https://doi.org/10.1007/JHEP02(2018)043} {\bibfield  {journal} {\bibinfo
  {journal} {JHEP}\ }\textbf {\bibinfo {volume} {02}},\ \bibinfo {pages}
  {043}},\ \Eprint {https://arxiv.org/abs/1712.00588} {arXiv:1712.00588
  [hep-ph]} \BibitemShut {NoStop}%
\bibitem [{\citenamefont {Beraudo}\ \emph {et~al.}(2022)\citenamefont
  {Beraudo}, \citenamefont {De~Pace}, \citenamefont {Monteno}, \citenamefont
  {Nardi},\ and\ \citenamefont {Prino}}]{Beraudo:2022dpz}%
  \BibitemOpen
  \bibfield  {author} {\bibinfo {author} {\bibfnamefont {A.}~\bibnamefont
  {Beraudo}}, \bibinfo {author} {\bibfnamefont {A.}~\bibnamefont {De~Pace}},
  \bibinfo {author} {\bibfnamefont {M.}~\bibnamefont {Monteno}}, \bibinfo
  {author} {\bibfnamefont {M.}~\bibnamefont {Nardi}},\ and\ \bibinfo {author}
  {\bibfnamefont {F.}~\bibnamefont {Prino}},\ }\href
  {https://doi.org/10.1140/epjc/s10052-022-10482-y} {\bibfield  {journal}
  {\bibinfo  {journal} {Eur. Phys. J. C}\ }\textbf {\bibinfo {volume} {82}},\
  \bibinfo {pages} {607} (\bibinfo {year} {2022})},\ \Eprint
  {https://arxiv.org/abs/2202.08732} {arXiv:2202.08732 [hep-ph]} \BibitemShut
  {NoStop}%
\bibitem [{\citenamefont {Heinz}\ and\ \citenamefont
  {Snellings}(2013)}]{Heinz:2013th}%
  \BibitemOpen
  \bibfield  {author} {\bibinfo {author} {\bibfnamefont {U.}~\bibnamefont
  {Heinz}}\ and\ \bibinfo {author} {\bibfnamefont {R.}~\bibnamefont
  {Snellings}},\ }\href {https://doi.org/10.1146/annurev-nucl-102212-170540}
  {\bibfield  {journal} {\bibinfo  {journal} {Ann. Rev. Nucl. Part. Sci.}\
  }\textbf {\bibinfo {volume} {63}},\ \bibinfo {pages} {123} (\bibinfo {year}
  {2013})},\ \Eprint {https://arxiv.org/abs/1301.2826} {arXiv:1301.2826
  [nucl-th]} \BibitemShut {NoStop}%
\bibitem [{\citenamefont {Gale}\ \emph {et~al.}(2013)\citenamefont {Gale},
  \citenamefont {Jeon},\ and\ \citenamefont {Schenke}}]{Gale:2013da}%
  \BibitemOpen
  \bibfield  {author} {\bibinfo {author} {\bibfnamefont {C.}~\bibnamefont
  {Gale}}, \bibinfo {author} {\bibfnamefont {S.}~\bibnamefont {Jeon}},\ and\
  \bibinfo {author} {\bibfnamefont {B.}~\bibnamefont {Schenke}},\ }\href
  {https://doi.org/10.1142/S0217751X13400113} {\bibfield  {journal} {\bibinfo
  {journal} {Int. J. Mod. Phys. A}\ }\textbf {\bibinfo {volume} {28}},\
  \bibinfo {pages} {1340011} (\bibinfo {year} {2013})},\ \Eprint
  {https://arxiv.org/abs/1301.5893} {arXiv:1301.5893 [nucl-th]} \BibitemShut
  {NoStop}%
\bibitem [{\citenamefont {Weller}\ and\ \citenamefont
  {Romatschke}(2017)}]{Weller:2017tsr}%
  \BibitemOpen
  \bibfield  {author} {\bibinfo {author} {\bibfnamefont {R.~D.}\ \bibnamefont
  {Weller}}\ and\ \bibinfo {author} {\bibfnamefont {P.}~\bibnamefont
  {Romatschke}},\ }\href {https://doi.org/10.1016/j.physletb.2017.09.077}
  {\bibfield  {journal} {\bibinfo  {journal} {Phys. Lett. B}\ }\textbf
  {\bibinfo {volume} {774}},\ \bibinfo {pages} {351} (\bibinfo {year}
  {2017})},\ \Eprint {https://arxiv.org/abs/1701.07145} {arXiv:1701.07145
  [nucl-th]} \BibitemShut {NoStop}%
\bibitem [{\citenamefont {Moreland}\ \emph {et~al.}(2015)\citenamefont
  {Moreland}, \citenamefont {Bernhard},\ and\ \citenamefont
  {Bass}}]{Moreland:2014oya}%
  \BibitemOpen
  \bibfield  {author} {\bibinfo {author} {\bibfnamefont {J.~S.}\ \bibnamefont
  {Moreland}}, \bibinfo {author} {\bibfnamefont {J.~E.}\ \bibnamefont
  {Bernhard}},\ and\ \bibinfo {author} {\bibfnamefont {S.~A.}\ \bibnamefont
  {Bass}},\ }\href {https://doi.org/10.1103/PhysRevC.92.011901} {\bibfield
  {journal} {\bibinfo  {journal} {Phys. Rev. C}\ }\textbf {\bibinfo {volume}
  {92}},\ \bibinfo {pages} {011901} (\bibinfo {year} {2015})},\ \Eprint
  {https://arxiv.org/abs/1412.4708} {arXiv:1412.4708 [nucl-th]} \BibitemShut
  {NoStop}%
\bibitem [{\citenamefont {Schenke}\ \emph {et~al.}(2012)\citenamefont
  {Schenke}, \citenamefont {Tribedy},\ and\ \citenamefont
  {Venugopalan}}]{Schenke:2012wb}%
  \BibitemOpen
  \bibfield  {author} {\bibinfo {author} {\bibfnamefont {B.}~\bibnamefont
  {Schenke}}, \bibinfo {author} {\bibfnamefont {P.}~\bibnamefont {Tribedy}},\
  and\ \bibinfo {author} {\bibfnamefont {R.}~\bibnamefont {Venugopalan}},\
  }\href {https://doi.org/10.1103/PhysRevLett.108.252301} {\bibfield  {journal}
  {\bibinfo  {journal} {Phys. Rev. Lett.}\ }\textbf {\bibinfo {volume} {108}},\
  \bibinfo {pages} {252301} (\bibinfo {year} {2012})},\ \Eprint
  {https://arxiv.org/abs/1202.6646} {arXiv:1202.6646 [nucl-th]} \BibitemShut
  {NoStop}%
\bibitem [{Note1()}]{Note1}%
  \BibitemOpen
  \bibinfo {note} {All these parameters were calibrated using Bayesian
  inference~\cite {Moreland:2018gsh}, but we have had to modify $k$ from 0.19
  to 0.3 in order to describe the KNO scaling in pp collisions measured in
  experiments, see right panel of Fig.~\ref {fig:initial}}\BibitemShut
  {NoStop}%
\bibitem [{\citenamefont {Schenke}\ \emph {et~al.}(2010)\citenamefont
  {Schenke}, \citenamefont {Jeon},\ and\ \citenamefont
  {Gale}}]{Schenke:2010nt}%
  \BibitemOpen
  \bibfield  {author} {\bibinfo {author} {\bibfnamefont {B.}~\bibnamefont
  {Schenke}}, \bibinfo {author} {\bibfnamefont {S.}~\bibnamefont {Jeon}},\ and\
  \bibinfo {author} {\bibfnamefont {C.}~\bibnamefont {Gale}},\ }\href
  {https://doi.org/10.1103/PhysRevC.82.014903} {\bibfield  {journal} {\bibinfo
  {journal} {Phys. Rev. C}\ }\textbf {\bibinfo {volume} {82}},\ \bibinfo
  {pages} {014903} (\bibinfo {year} {2010})},\ \Eprint
  {https://arxiv.org/abs/1004.1408} {arXiv:1004.1408 [hep-ph]} \BibitemShut
  {NoStop}%
\bibitem [{\citenamefont {Schenke}\ \emph {et~al.}(2011)\citenamefont
  {Schenke}, \citenamefont {Jeon},\ and\ \citenamefont
  {Gale}}]{Schenke:2010rr}%
  \BibitemOpen
  \bibfield  {author} {\bibinfo {author} {\bibfnamefont {B.}~\bibnamefont
  {Schenke}}, \bibinfo {author} {\bibfnamefont {S.}~\bibnamefont {Jeon}},\ and\
  \bibinfo {author} {\bibfnamefont {C.}~\bibnamefont {Gale}},\ }\href
  {https://doi.org/10.1103/PhysRevLett.106.042301} {\bibfield  {journal}
  {\bibinfo  {journal} {Phys. Rev. Lett.}\ }\textbf {\bibinfo {volume} {106}},\
  \bibinfo {pages} {042301} (\bibinfo {year} {2011})},\ \Eprint
  {https://arxiv.org/abs/1009.3244} {arXiv:1009.3244 [hep-ph]} \BibitemShut
  {NoStop}%
\bibitem [{\citenamefont {Paquet}\ \emph {et~al.}(2016)\citenamefont {Paquet},
  \citenamefont {Shen}, \citenamefont {Denicol}, \citenamefont {Luzum},
  \citenamefont {Schenke}, \citenamefont {Jeon},\ and\ \citenamefont
  {Gale}}]{Paquet:2015lta}%
  \BibitemOpen
  \bibfield  {author} {\bibinfo {author} {\bibfnamefont {J.-F.}\ \bibnamefont
  {Paquet}}, \bibinfo {author} {\bibfnamefont {C.}~\bibnamefont {Shen}},
  \bibinfo {author} {\bibfnamefont {G.~S.}\ \bibnamefont {Denicol}}, \bibinfo
  {author} {\bibfnamefont {M.}~\bibnamefont {Luzum}}, \bibinfo {author}
  {\bibfnamefont {B.}~\bibnamefont {Schenke}}, \bibinfo {author} {\bibfnamefont
  {S.}~\bibnamefont {Jeon}},\ and\ \bibinfo {author} {\bibfnamefont
  {C.}~\bibnamefont {Gale}},\ }\href
  {https://doi.org/10.1103/PhysRevC.93.044906} {\bibfield  {journal} {\bibinfo
  {journal} {Phys. Rev. C}\ }\textbf {\bibinfo {volume} {93}},\ \bibinfo
  {pages} {044906} (\bibinfo {year} {2016})},\ \Eprint
  {https://arxiv.org/abs/1509.06738} {arXiv:1509.06738 [hep-ph]} \BibitemShut
  {NoStop}%
\bibitem [{\citenamefont {Bazavov}\ \emph {et~al.}(2014)\citenamefont {Bazavov}
  \emph {et~al.}}]{HotQCD:2014kol}%
  \BibitemOpen
  \bibfield  {author} {\bibinfo {author} {\bibfnamefont {A.}~\bibnamefont
  {Bazavov}} \emph {et~al.} (\bibinfo {collaboration} {HotQCD}),\ }\href
  {https://doi.org/10.1103/PhysRevD.90.094503} {\bibfield  {journal} {\bibinfo
  {journal} {Phys. Rev. D}\ }\textbf {\bibinfo {volume} {90}},\ \bibinfo
  {pages} {094503} (\bibinfo {year} {2014})},\ \Eprint
  {https://arxiv.org/abs/1407.6387} {arXiv:1407.6387 [hep-lat]} \BibitemShut
  {NoStop}%
\bibitem [{\citenamefont {Denicol}\ \emph {et~al.}(2009)\citenamefont
  {Denicol}, \citenamefont {Kodama}, \citenamefont {Koide},\ and\ \citenamefont
  {Mota}}]{Denicol:2009am}%
  \BibitemOpen
  \bibfield  {author} {\bibinfo {author} {\bibfnamefont {G.~S.}\ \bibnamefont
  {Denicol}}, \bibinfo {author} {\bibfnamefont {T.}~\bibnamefont {Kodama}},
  \bibinfo {author} {\bibfnamefont {T.}~\bibnamefont {Koide}},\ and\ \bibinfo
  {author} {\bibfnamefont {P.}~\bibnamefont {Mota}},\ }\href
  {https://doi.org/10.1103/PhysRevC.80.064901} {\bibfield  {journal} {\bibinfo
  {journal} {Phys. Rev. C}\ }\textbf {\bibinfo {volume} {80}},\ \bibinfo
  {pages} {064901} (\bibinfo {year} {2009})},\ \Eprint
  {https://arxiv.org/abs/0903.3595} {arXiv:0903.3595 [hep-ph]} \BibitemShut
  {NoStop}%
\bibitem [{\citenamefont {Cooper}\ and\ \citenamefont
  {Frye}(1974)}]{Cooper:1974mv}%
  \BibitemOpen
  \bibfield  {author} {\bibinfo {author} {\bibfnamefont {F.}~\bibnamefont
  {Cooper}}\ and\ \bibinfo {author} {\bibfnamefont {G.}~\bibnamefont {Frye}},\
  }\href {https://doi.org/10.1103/PhysRevD.10.186} {\bibfield  {journal}
  {\bibinfo  {journal} {Phys. Rev. D}\ }\textbf {\bibinfo {volume} {10}},\
  \bibinfo {pages} {186} (\bibinfo {year} {1974})}\BibitemShut {NoStop}%
\bibitem [{\citenamefont {Hanus}\ \emph {et~al.}(2019)\citenamefont {Hanus},
  \citenamefont {Mazeliauskas},\ and\ \citenamefont {Reygers}}]{Hanus:2019fnc}%
  \BibitemOpen
  \bibfield  {author} {\bibinfo {author} {\bibfnamefont {P.}~\bibnamefont
  {Hanus}}, \bibinfo {author} {\bibfnamefont {A.}~\bibnamefont
  {Mazeliauskas}},\ and\ \bibinfo {author} {\bibfnamefont {K.}~\bibnamefont
  {Reygers}},\ }\href {https://doi.org/10.1103/PhysRevC.100.064903} {\bibfield
  {journal} {\bibinfo  {journal} {Phys. Rev. C}\ }\textbf {\bibinfo {volume}
  {100}},\ \bibinfo {pages} {064903} (\bibinfo {year} {2019})},\ \Eprint
  {https://arxiv.org/abs/1908.02792} {arXiv:1908.02792 [hep-ph]} \BibitemShut
  {NoStop}%
\bibitem [{\citenamefont {Adam}\ \emph
  {et~al.}(2017{\natexlab{a}})\citenamefont {Adam} \emph
  {et~al.}}]{ALICE:2015olq}%
  \BibitemOpen
  \bibfield  {author} {\bibinfo {author} {\bibfnamefont {J.}~\bibnamefont
  {Adam}} \emph {et~al.} (\bibinfo {collaboration} {ALICE}),\ }\href
  {https://doi.org/10.1140/epjc/s10052-016-4571-1} {\bibfield  {journal}
  {\bibinfo  {journal} {Eur. Phys. J. C}\ }\textbf {\bibinfo {volume} {77}},\
  \bibinfo {pages} {33} (\bibinfo {year} {2017}{\natexlab{a}})},\ \Eprint
  {https://arxiv.org/abs/1509.07541} {arXiv:1509.07541 [nucl-ex]} \BibitemShut
  {NoStop}%
\bibitem [{\citenamefont {Koba}\ \emph {et~al.}(1972)\citenamefont {Koba},
  \citenamefont {Nielsen},\ and\ \citenamefont {Olesen}}]{Koba:1972ng}%
  \BibitemOpen
  \bibfield  {author} {\bibinfo {author} {\bibfnamefont {Z.}~\bibnamefont
  {Koba}}, \bibinfo {author} {\bibfnamefont {H.~B.}\ \bibnamefont {Nielsen}},\
  and\ \bibinfo {author} {\bibfnamefont {P.}~\bibnamefont {Olesen}},\ }\href
  {https://doi.org/10.1016/0550-3213(72)90551-2} {\bibfield  {journal}
  {\bibinfo  {journal} {Nucl. Phys. B}\ }\textbf {\bibinfo {volume} {40}},\
  \bibinfo {pages} {317} (\bibinfo {year} {1972})}\BibitemShut {NoStop}%
\bibitem [{\citenamefont {Alioli}\ \emph {et~al.}(2010)\citenamefont {Alioli},
  \citenamefont {Nason}, \citenamefont {Oleari},\ and\ \citenamefont
  {Re}}]{Alioli:2010xd}%
  \BibitemOpen
  \bibfield  {author} {\bibinfo {author} {\bibfnamefont {S.}~\bibnamefont
  {Alioli}}, \bibinfo {author} {\bibfnamefont {P.}~\bibnamefont {Nason}},
  \bibinfo {author} {\bibfnamefont {C.}~\bibnamefont {Oleari}},\ and\ \bibinfo
  {author} {\bibfnamefont {E.}~\bibnamefont {Re}},\ }\href
  {https://doi.org/10.1007/JHEP06(2010)043} {\bibfield  {journal} {\bibinfo
  {journal} {JHEP}\ }\textbf {\bibinfo {volume} {06}},\ \bibinfo {pages}
  {043}},\ \Eprint {https://arxiv.org/abs/1002.2581} {arXiv:1002.2581 [hep-ph]}
  \BibitemShut {NoStop}%
\bibitem [{\citenamefont {Arnison}\ \emph {et~al.}(1983)\citenamefont {Arnison}
  \emph {et~al.}}]{UA1:1983hhd}%
  \BibitemOpen
  \bibfield  {author} {\bibinfo {author} {\bibfnamefont {G.}~\bibnamefont
  {Arnison}} \emph {et~al.} (\bibinfo {collaboration} {UA1}),\ }\href
  {https://doi.org/10.1016/0370-2693(83)90254-X} {\bibfield  {journal}
  {\bibinfo  {journal} {Phys. Lett. B}\ }\textbf {\bibinfo {volume} {132}},\
  \bibinfo {pages} {214} (\bibinfo {year} {1983})}\BibitemShut {NoStop}%
\bibitem [{\citenamefont {Altenkort}\ \emph {et~al.}(2023)\citenamefont
  {Altenkort}, \citenamefont {Kaczmarek}, \citenamefont {Larsen}, \citenamefont
  {Mukherjee}, \citenamefont {Petreczky}, \citenamefont {Shu},\ and\
  \citenamefont {Stendebach}}]{Altenkort:2023oms}%
  \BibitemOpen
  \bibfield  {author} {\bibinfo {author} {\bibfnamefont {L.}~\bibnamefont
  {Altenkort}}, \bibinfo {author} {\bibfnamefont {O.}~\bibnamefont
  {Kaczmarek}}, \bibinfo {author} {\bibfnamefont {R.}~\bibnamefont {Larsen}},
  \bibinfo {author} {\bibfnamefont {S.}~\bibnamefont {Mukherjee}}, \bibinfo
  {author} {\bibfnamefont {P.}~\bibnamefont {Petreczky}}, \bibinfo {author}
  {\bibfnamefont {H.-T.}\ \bibnamefont {Shu}},\ and\ \bibinfo {author}
  {\bibfnamefont {S.}~\bibnamefont {Stendebach}},\ }\href@noop {} {\  (\bibinfo
  {year} {2023})},\ \Eprint {https://arxiv.org/abs/2302.08501}
  {arXiv:2302.08501 [hep-lat]} \BibitemShut {NoStop}%
\bibitem [{\citenamefont {Acharya}\ \emph
  {et~al.}(2021{\natexlab{c}})\citenamefont {Acharya} \emph
  {et~al.}}]{ALICE:2021mgk}%
  \BibitemOpen
  \bibfield  {author} {\bibinfo {author} {\bibfnamefont {S.}~\bibnamefont
  {Acharya}} \emph {et~al.} (\bibinfo {collaboration} {ALICE}),\ }\href
  {https://doi.org/10.1007/JHEP05(2021)220} {\bibfield  {journal} {\bibinfo
  {journal} {JHEP}\ }\textbf {\bibinfo {volume} {05}},\ \bibinfo {pages}
  {220}},\ \Eprint {https://arxiv.org/abs/2102.13601} {arXiv:2102.13601
  [nucl-ex]} \BibitemShut {NoStop}%
\bibitem [{\citenamefont {Sjostrand}\ \emph {et~al.}(2006)\citenamefont
  {Sjostrand}, \citenamefont {Mrenna},\ and\ \citenamefont
  {Skands}}]{Sjostrand:2006za}%
  \BibitemOpen
  \bibfield  {author} {\bibinfo {author} {\bibfnamefont {T.}~\bibnamefont
  {Sjostrand}}, \bibinfo {author} {\bibfnamefont {S.}~\bibnamefont {Mrenna}},\
  and\ \bibinfo {author} {\bibfnamefont {P.~Z.}\ \bibnamefont {Skands}},\
  }\href {https://doi.org/10.1088/1126-6708/2006/05/026} {\bibfield  {journal}
  {\bibinfo  {journal} {JHEP}\ }\textbf {\bibinfo {volume} {05}},\ \bibinfo
  {pages} {026}},\ \Eprint {https://arxiv.org/abs/hep-ph/0603175}
  {arXiv:hep-ph/0603175} \BibitemShut {NoStop}%
\bibitem [{\citenamefont {Adam}\ \emph
  {et~al.}(2017{\natexlab{b}})\citenamefont {Adam} \emph
  {et~al.}}]{ALICE:2016fzo}%
  \BibitemOpen
  \bibfield  {author} {\bibinfo {author} {\bibfnamefont {J.}~\bibnamefont
  {Adam}} \emph {et~al.} (\bibinfo {collaboration} {ALICE}),\ }\href
  {https://doi.org/10.1038/nphys4111} {\bibfield  {journal} {\bibinfo
  {journal} {Nature Phys.}\ }\textbf {\bibinfo {volume} {13}},\ \bibinfo
  {pages} {535} (\bibinfo {year} {2017}{\natexlab{b}})},\ \Eprint
  {https://arxiv.org/abs/1606.07424} {arXiv:1606.07424 [nucl-ex]} \BibitemShut
  {NoStop}%
\bibitem [{\citenamefont {Webber}(1984)}]{Webber:1983if}%
  \BibitemOpen
  \bibfield  {author} {\bibinfo {author} {\bibfnamefont {B.~R.}\ \bibnamefont
  {Webber}},\ }\href {https://doi.org/10.1016/0550-3213(84)90333-X} {\bibfield
  {journal} {\bibinfo  {journal} {Nucl. Phys. B}\ }\textbf {\bibinfo {volume}
  {238}},\ \bibinfo {pages} {492} (\bibinfo {year} {1984})}\BibitemShut
  {NoStop}%
\bibitem [{\citenamefont {Acharya}\ \emph
  {et~al.}(2022{\natexlab{a}})\citenamefont {Acharya} \emph
  {et~al.}}]{ALICE:2021dhb}%
  \BibitemOpen
  \bibfield  {author} {\bibinfo {author} {\bibfnamefont {S.}~\bibnamefont
  {Acharya}} \emph {et~al.} (\bibinfo {collaboration} {ALICE}),\ }\href
  {https://doi.org/10.1103/PhysRevD.105.L011103} {\bibfield  {journal}
  {\bibinfo  {journal} {Phys. Rev. D}\ }\textbf {\bibinfo {volume} {105}},\
  \bibinfo {pages} {L011103} (\bibinfo {year} {2022}{\natexlab{a}})},\ \Eprint
  {https://arxiv.org/abs/2105.06335} {arXiv:2105.06335 [nucl-ex]} \BibitemShut
  {NoStop}%
\bibitem [{\citenamefont {Acharya}\ \emph
  {et~al.}(2022{\natexlab{b}})\citenamefont {Acharya} \emph
  {et~al.}}]{ALICE:2021rxa}%
  \BibitemOpen
  \bibfield  {author} {\bibinfo {author} {\bibfnamefont {S.}~\bibnamefont
  {Acharya}} \emph {et~al.} (\bibinfo {collaboration} {ALICE}),\ }\href
  {https://doi.org/10.1007/JHEP01(2022)174} {\bibfield  {journal} {\bibinfo
  {journal} {JHEP}\ }\textbf {\bibinfo {volume} {01}},\ \bibinfo {pages}
  {174}},\ \Eprint {https://arxiv.org/abs/2110.09420} {arXiv:2110.09420
  [nucl-ex]} \BibitemShut {NoStop}%
\bibitem [{\citenamefont {Acharya}\ \emph {et~al.}(2023)\citenamefont {Acharya}
  \emph {et~al.}}]{ALICE:2021bib}%
  \BibitemOpen
  \bibfield  {author} {\bibinfo {author} {\bibfnamefont {S.}~\bibnamefont
  {Acharya}} \emph {et~al.} (\bibinfo {collaboration} {ALICE}),\ }\href
  {https://doi.org/10.1016/j.physletb.2023.137796} {\bibfield  {journal}
  {\bibinfo  {journal} {Phys. Lett. B}\ }\textbf {\bibinfo {volume} {839}},\
  \bibinfo {pages} {137796} (\bibinfo {year} {2023})},\ \Eprint
  {https://arxiv.org/abs/2112.08156} {arXiv:2112.08156 [nucl-ex]} \BibitemShut
  {NoStop}%
\bibitem [{\citenamefont {Sirunyan}\ \emph {et~al.}(2021)\citenamefont
  {Sirunyan} \emph {et~al.}}]{CMS:2020qul}%
  \BibitemOpen
  \bibfield  {author} {\bibinfo {author} {\bibfnamefont {A.~M.}\ \bibnamefont
  {Sirunyan}} \emph {et~al.} (\bibinfo {collaboration} {CMS}),\ }\href
  {https://doi.org/10.1016/j.physletb.2020.136036} {\bibfield  {journal}
  {\bibinfo  {journal} {Phys. Lett. B}\ }\textbf {\bibinfo {volume} {813}},\
  \bibinfo {pages} {136036} (\bibinfo {year} {2021})},\ \Eprint
  {https://arxiv.org/abs/2009.07065} {arXiv:2009.07065 [hep-ex]} \BibitemShut
  {NoStop}%
\bibitem [{\citenamefont {Qiu}\ and\ \citenamefont {Heinz}(2011)}]{Qiu:2011iv}%
  \BibitemOpen
  \bibfield  {author} {\bibinfo {author} {\bibfnamefont {Z.}~\bibnamefont
  {Qiu}}\ and\ \bibinfo {author} {\bibfnamefont {U.~W.}\ \bibnamefont
  {Heinz}},\ }\href {https://doi.org/10.1103/PhysRevC.84.024911} {\bibfield
  {journal} {\bibinfo  {journal} {Phys. Rev. C}\ }\textbf {\bibinfo {volume}
  {84}},\ \bibinfo {pages} {024911} (\bibinfo {year} {2011})},\ \Eprint
  {https://arxiv.org/abs/1104.0650} {arXiv:1104.0650 [nucl-th]} \BibitemShut
  {NoStop}%
\bibitem [{\citenamefont {Moreland}\ \emph {et~al.}(2020)\citenamefont
  {Moreland}, \citenamefont {Bernhard},\ and\ \citenamefont
  {Bass}}]{Moreland:2018gsh}%
  \BibitemOpen
  \bibfield  {author} {\bibinfo {author} {\bibfnamefont {J.~S.}\ \bibnamefont
  {Moreland}}, \bibinfo {author} {\bibfnamefont {J.~E.}\ \bibnamefont
  {Bernhard}},\ and\ \bibinfo {author} {\bibfnamefont {S.~A.}\ \bibnamefont
  {Bass}},\ }\href {https://doi.org/10.1103/PhysRevC.101.024911} {\bibfield
  {journal} {\bibinfo  {journal} {Phys. Rev. C}\ }\textbf {\bibinfo {volume}
  {101}},\ \bibinfo {pages} {024911} (\bibinfo {year} {2020})},\ \Eprint
  {https://arxiv.org/abs/1808.02106} {arXiv:1808.02106 [nucl-th]} \BibitemShut
  {NoStop}%
\end{thebibliography}%

\end{document}